\documentclass[3p]{elsarticle}

\usepackage{hyperref}

\usepackage{amssymb}
\usepackage{bm}
\usepackage{dcolumn}
\usepackage{amssymb}
\usepackage{amsmath}
\usepackage{graphicx}

\usepackage{epstopdf}
\usepackage{epsfig}
\usepackage{xcolor}
\usepackage{color}
\usepackage{subfig}
\usepackage{booktabs}
\usepackage{multirow}
\usepackage{setspace}
\usepackage{array}
\usepackage{threeparttable}
\usepackage{txfonts}

\usepackage{exscale}
\usepackage{relsize}

\begin{document}

\journal{
PLB, published in \href{https://doi.org/10.1016/j.physletb.2024.138951}{Phys.~Lett.~B 856 (2024) 138951. }}

\begin{frontmatter}



  \newcommand*{\PKU}{School of Physics, Peking University, Beijing 100871,
    China}
  \newcommand*{\CHEP}{Center for High Energy Physics, Peking University, Beijing 100871, China}

\title{Energy-dependent intrinsic time delay of gamma-ray bursts on testing Lorentz invariance violation}

  \author[a]{Hanlin Song}
  \author[a,b]{Bo-Qiang Ma\corref{cor1}}

  \address[a]{\PKU}
  \address[b]{\CHEP}

  \cortext[cor1]{Corresponding author \ead{mabq@pku.edu.cn}}

\begin{abstract}
High-energy photons of gamma-ray bursts (GRBs) might be emitted at different intrinsic times with energy dependence at the source. In this letter, we expand the model from previous works on testing the Lorentz Invariance Violation (LV) with the observed GRB data from the Fermi Gamma-ray Space Telescope. We reanalyze the previous data with the full Bayesian parameter estimation method and get consistent results by assuming that the time delays are due to an LV term and a constant intrinsic time delay term. Subsequently, we neglect the LV effect and only consider the intrinsic time delay effect. We assume a common intrinsic time delay term along with a source energy correlated time delay of high-energy photons. We find that the energy-dependent emission times can also explain the observed GRB data of high-energy photon events. Finally, we integrate these two physical mechanisms into a unified model to distinguish and evaluate their respective contributions using the observed GRB data.
\end{abstract}

\end{frontmatter}

\section{Introduction}
Lorentz invariance is a key assumption in many aspects of modern physics. However, 
to achieve the goal of unification of the standard model 
with general relativity, 
Lorentz invariance violation (LV) is predicted in many theories, 
such as string theory \cite{Amelino-Camelia:1996bln, Amelino-Camelia:1997ieq, Ellis:1999rz, Ellis:1999uh, Ellis:2008gg, Li:2009tt, Li:2021gah, Li:2021eza}, 
loop quantum gravity \cite{Gambini:1998it, Alfaro:1999wd,Li:2022szn}, 
and doubly-special relativity \cite{Amelino-Camelia:2002cqb, Amelino-Camelia:2002uql, Amelino-Camelia:2000stu}. 
In most of these theories, the LV might happen around the Planck scale, $E_{\rm P} \equiv \sqrt{\hbar c^5/G} \simeq 1.22 \times 10^{19}$ GeV. 
Consequently, 
for a  photon with energy $E \ll E_{\rm P
}$ and momentum $p$, the dispersion relation for photons can be modified by the leading terms of Taylor series \cite{Jacob:2008bw,Xiao:2009xe}:

\begin{equation}
E^2=p^2c^2\left[1-s_n\left(\frac{pc}{E_{\mathrm{LV},n}}\right)^n\right],
\end{equation}
where $s_n \equiv \pm 1$ is the indicator for high-energy photons traveling faster ($s_n = - 1$) or slower ($s_n = + 1$) than the low-energy photons and $E_{{\rm LV},n}$ denotes the $n$-th order energy scale of LV to be determined by the observation. 
By applying the relation $v = \partial E / \partial p$, the modified velocity relation is: 

\begin{equation}
v(E)=c\left[1-s_n\frac{n+1}{2}\left(\frac{pc}{E_{\mathrm{LV},n}}\right)^n\right].
\end{equation}

Because of the great suppression of $E_{{\rm LV},n}$ around the Planck scale, the LV is hardly to be 
observable by experiments 
on the Earth. 
Amelino-Camelia {\it et al.}~\cite{Amelino-Camelia:1996bln, Amelino-Camelia:1997ieq} first suggested using the time delay between high-energy photons and low-energy photons of Gamma-Ray Bursts (GRBs), whereas the tiny velocity difference can be accumulated into observable arrival time discrepancy due to the long cosmological journey of these photons to arrive at the Earth, to enable the measurement of LV effect. 
By taking into account the expansion of the universe, the time delay caused by LV can be expressed as \cite{Jacob:2008bw,Zhu:2022blp}:
\begin{equation}
\label{lorentzdelay}
\Delta t_{\mathrm{LV}}=s_{n}\frac{1+n}{2H_{0}}\frac{E_{\mathrm{h}}^{n}-E_{1}^{n}}{E_{\mathrm{LV},n}^{n}}\int_{0}^{z}\frac{(1+z')^{n}\mathrm{d}z'}{\sqrt{\Omega_{\mathrm{m}}(1+z')^{3}+\Omega_{\Lambda}}},
\end{equation}
where $E_{\rm h}$ and $E_{\rm l}$ are energies of high- and low-energy photons, $z$ is the redshift of GRB, $H_0$ is the Hubble constant, and $\Omega_{\mathrm{m}}$ and $\Omega_{\Lambda}$ are matter density parameter and dark energy density parameter of the $\Lambda {\rm CDM}$ model.

The observed time delay $\Delta t_{\mathrm{obs}}$ of some GRB photon events with energies ranging from 33.6 GeV to 146.6 GeV at the source frame can be explained by the LV time delay term in Eq.~\ref{lorentzdelay} and a common intrinsic time delay term \cite{Ellis:2005sjy, Shao:2009bv, Zhang:2014wpb, Xu:2016zxi, Xu:2016zsa, Zhu:2021pml, Zhu:2021wtw, Zhu:2022usw, Huang:2019etr}. Due to the expansion of the universe, it can be expressed as:
\begin{equation}
\label{obsdelay}
\Delta t_{\mathrm{obs}}=\Delta t_{\mathrm{LV}}+(1+z)\Delta t_{\mathrm{in}},
\end{equation}
where $\Delta t_{\mathrm{in}}$ is the intrinsic emission time delay between high-energy photon and low-energy photon in the GRB source frame. 

The GRB data can be obtained from the Fermi Gamma-ray Space Telescope (FGST), which is comprised of the Fermi Large Area Telescope (LAT) for observing high-energy photons and the Gamma-ray Burst Monitor (GBM) for detecting low-energy photons \cite{Fermi-LAT:2009ihh, Meegan:2009qu}. 
Then $\Delta t_{\rm LV}$ can be calculated with given $z$, $E_{{\rm h}}$, $E_{{\rm l}}$ of GRB events from FGST together with $E_{\mathrm{LV},n}$ as a fitting parameter for a fixed $n$.
The observed time delay $\Delta t_{\mathrm{obs}}$ is:
\begin{equation}
    \Delta t_{\mathrm{obs}} = t_{\rm high} - t_{\rm low},
\end{equation}
where $t_{\rm high}$ is the observed arrival time of high-energy photon and $t_{\rm low}$ is the time of first significant peak of low-energy photon. As shown in previous works~\cite{Xu:2016zxi, Xu:2016zsa}, after analyzing high-energy photon GRB  events detected by FGST,  the results suggest that $n=  1$, $s_n = + 1$, $E_{\rm LV,1} \simeq 3.60 \times 10^{17}$ GeV, and $\Delta t_{\rm in} = -10.7$ s for the mainline photons (see Fig.~2 of \cite{Xu:2016zsa}). These findings suggest that only the subluminal aspect of cosmic photon Lorentz violation remains permissible. This conclusion aligns with the stringent constraints on superluminal Lorentz violation of cosmic photons derived from vacuum birefringence (see Section 4.2 of the review \cite{He:2022gyk}) and the detection of PeV photons by LHAASO \cite{LHAASO:2021gok,Li:2021tcw,He:2022jdl,He:2023ydr}. Furthermore, it is in agreement with the theoretical predictions of string theory \cite{Li:2021gah,Li:2021eza,Li:2021tcw} and loop quantum gravity \cite{Li:2022szn}. 

In principle, other physical mechanisms could also explain the observed time delay. For example, high-energy photons with different values might be emitted at different times at the source. The detailed model of intrinsic time delay should be considered. Therefore, in this work, we explore another potential mechanism for observed time delay in which we assume a correlation between the energy value of photons and intrinsic emission time. We also combine the two physical mechanisms together and try to distinguish them using the same FGST GRB data as \cite{Xu:2016zxi, Xu:2016zsa}. This letter is arranged as follows. In Sec.~\ref{Methods}, we introduce the full Bayesian method for the parameter estimation. In Sec.~\ref{results}, we introduce three models for analyzing the LV effects and intrinsic time delay.  In Sec.~\ref{summary}, we give a summary of our results.

\section{Parameter Estimation Method}
\label{Methods}
In this section, we introduce a full Bayesian parameter estimation method for a general multiple linear model.  

Consider the multiple linear model for the true value of physical quantity $y_t$, which depends on 
 $n$ physical quantities with true values $\{x_{t1}, \cdots, x_{tn}\}$:

\begin{equation}
\begin{aligned}
     y_t = f(x_{t1}, \cdots , x_{tn}) = \beta_0 + \beta_1x_{t1} + \cdots + \beta_nx_{tn}.
\end{aligned}
\end{equation}
In reality, there will always be a measurement precision limit  because of the existence of noise, which leads to measurement errors:

\begin{equation}
    \begin{cases}
        y_m = y_t + y_e, \\
        x_{mi} = x_{ti} + x_{ei},
    \end{cases}
\end{equation}
where the subscript $mi$ denotes the measurement value, and the subscript $ei$ denotes the measurement error for the $i$-th variable respectively.

Assuming that these measurement error variables are independent and follow Gaussian distribution, then the likelihood function for measuring $p$ times of $y_{m}$, given the coefficients $\beta_0, \cdots, \beta_{n}$ and corresponding $\{\boldsymbol{x_{m}}\}$, can be written as:
\begin{equation}
    p(\{y_{m}\}|\{\boldsymbol{x_{m}}\},{\beta_0, \cdots, \beta_n}) = \prod_{j=1}^p \frac{1}{\sqrt{2\pi \left( \sigma_{x_{m1,j}}^2\beta_1^2 + , \cdots , + \sigma_{x_{mn,j}}^2\beta_n^2 + \sigma_{y_{m,j}}^2 \right) }}\exp{\left(-\frac{\left( y_{m,j}  -  \beta_0 - \beta_1x_{m1,j} \ - , \cdots , \ - \beta_nx_{mn,j}\right)^2}{2\left( \sigma_{x_{m1,j}}^2\beta_1^2 + , \cdots , + \sigma_{x_{mn,j}}^2\beta_n^2 + \sigma_{y_{m,j}}^2 \right)} \right)}, 
\end{equation}
where $\sigma_{x_{mi,j}}$ and $\sigma_{y_{m,j}}$ denote the uncertainty of variables $x_{mi}$ and $y_{m}$ during the $j$-th measurement.  

According to the Bayesian theorem, the posterior is:
\begin{equation}
    p({\beta_0, \cdots, \beta_n}|\{y_{m}\}, \{\boldsymbol{x_{m}}\}) \propto p(\{y_{m}\}|\{\boldsymbol{x_{m}}\},{\beta_0, \cdots, \beta_n}) p(\beta_0, \cdots, \beta_n),
\end{equation}
where $p(\beta_0, \cdots, \beta_n)$ is the prior for coefficients. Assuming all the coefficients are independent,  the prior can be written as:
\begin{equation}
    p(\beta_0, \cdots, \beta_n) = p(\beta_0) \cdots p(\beta_n).
\end{equation}

Given the model and measured data, we can obtain the results for parameter estimation using the above method.

\section{Models and Results}
\label{results}

In this section, we introduce three models to explain the observed time delay between high-energy and low-energy photons of GRBs. In Model A, we consider the LV time delay term and a common intrinsic time delay term. In Model B, we neglect the LV effects and consider the source energy correspondent intrinsic time delay term and a common intrinsic time delay term. In Model C, we combine the above two physical mechanisms as a unified model. We apply the same FGST GRB data from \cite{Xu:2016zxi, Xu:2016zsa} to fit the three models. A total of 10 GRBs with 14 high-energy photon events ranging from 33.6~GeV to 146.6~GeV at the source frame and within a 90~s time window are considered. The \texttt{bilby} \cite{Ashton:2018jfp, Romero-Shaw:2020owr} package is adopted in our calculation.

\subsection{Model A}

In Model A, we revisit the model mentioned in previous works \cite{Xu:2016zxi, Xu:2016zsa} where the observed time delay consists of two parts. One part comes from the LV time delay term as in Eq.~\ref{lorentzdelay}. The other part is a common intrinsic time delay term for GRBs. Considering $n=1$, $s_n = +1$, and neglecting $E_{\rm l}$, the model can be rewritten as:

\begin{equation}
\frac{\Delta t_{\mathrm{obs}}}{1+z}= \Delta t_{\mathrm{in}} + \frac{\Delta t_{\rm LV}}{1+z} = \Delta t_{\mathrm{in}} + a_{\rm LV}K_1,
\end{equation}
where $a_{\rm LV} = 1/E_{\rm LV}$ and $K_1$ is:
\begin{equation}
K_1=\frac{1}{H_0}\frac{E_\mathrm{h}}{1+z}\int_0^z\frac{(1+z')\mathrm{d}z'}{\sqrt{\Omega_\mathrm{m}(1+z')^3+\Omega_\Lambda}}.
\end{equation}

We assume that the common intrinsic time delay $\Delta t_{\mathrm{in}}$ follows a Gaussian distribution:
\begin{equation}
    p\left( \Delta t_{\mathrm{in}} \right) \sim \mathcal{N} \left(\mu, \sigma\right),
\end{equation}
where $\mu$ is the mean value of common intrinsic time delay $\Delta t_{\rm obs}$ and $\sigma$ is the standard variance.

Considering the energy and time observational uncertainty \cite{Fermi-LAT:2009ihh}, we take same value of $\sigma_{x_{m,j}}$ and take $\sigma_{y_{m,j}} = 0$, which are the same as the works in \cite{Xu:2016zxi, Xu:2016zsa, Zhu:2021pml}. Here $y_{m,j} = \Delta t_{\mathrm{obs},j}/{1+z_j}$ and $x_{m,j} = K_{1, j}$.
Marginalizing over the variable $\Delta t_{\mathrm{in}}$, the  posterior now becomes:
\begin{equation}
p \propto \exp\left[-\frac{1}{2}\sum_{j=1}^{p}\left(\frac{\left(\frac{\Delta t_{\mathrm{obs},j}}{1+z_{j}} -\mu -a_{\mathrm{LV}}K_{1,j}\right)^{2}}{\sigma^{2}+a_{\mathrm{LV}}^{2}\sigma_{K_{1,j}}^{2}}+\ln\left(\sigma^{2}+a_{\mathrm{LV}}^{2}\sigma_{K_{1,j}}^{2}\right)\right)\right]p\left(\mu\right)p\left(\sigma\right)p\left(a_{\rm LV}\right).
\end{equation}

We assume that these 3 parameters, $\mu$, $\sigma$ , and $a_{\rm LV}$, have flat priors:
\begin{align}
    \begin{cases}
        p\left(\mu\right) \sim U\left[-30, 30 \right] \ {\rm s}  , \\
        p\left(\sigma\right) \sim U\left[0, 30 \right] \ {\rm s}, \\
        p\left(a_{\rm LV}\right) \sim U\left[0, 30 \right] \times 10^{-18} \ \rm{GeV^{-1}}.
\end{cases}  
\end{align}

\begin{figure}[htbp]
    \centering
    \begin{minipage}{0.43\textwidth}
        \centering
        \includegraphics[width=\textwidth]{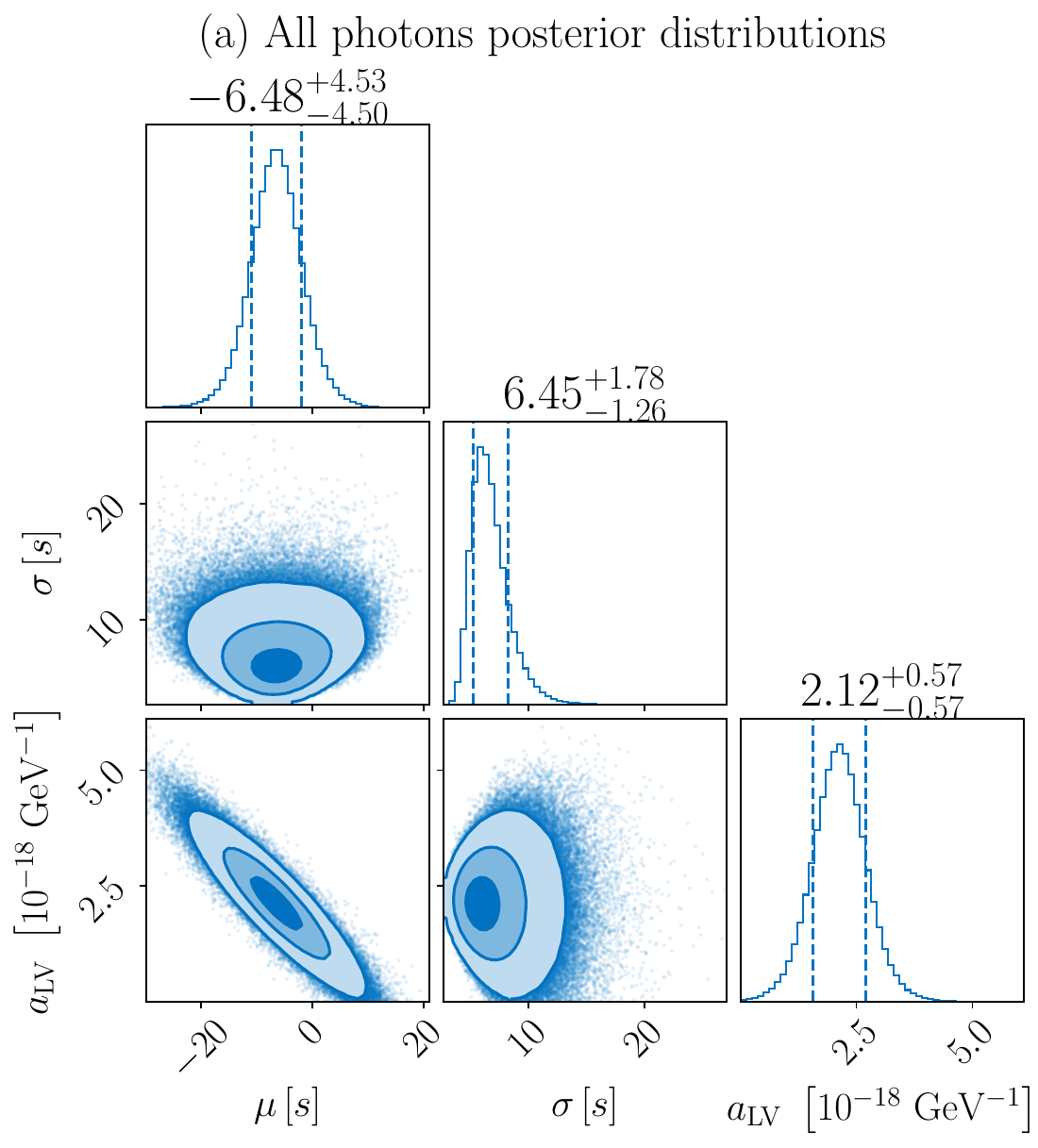}
    \end{minipage}
    \hfill
    \begin{minipage}{0.55\textwidth}
        \centering
        \includegraphics[width=\textwidth]{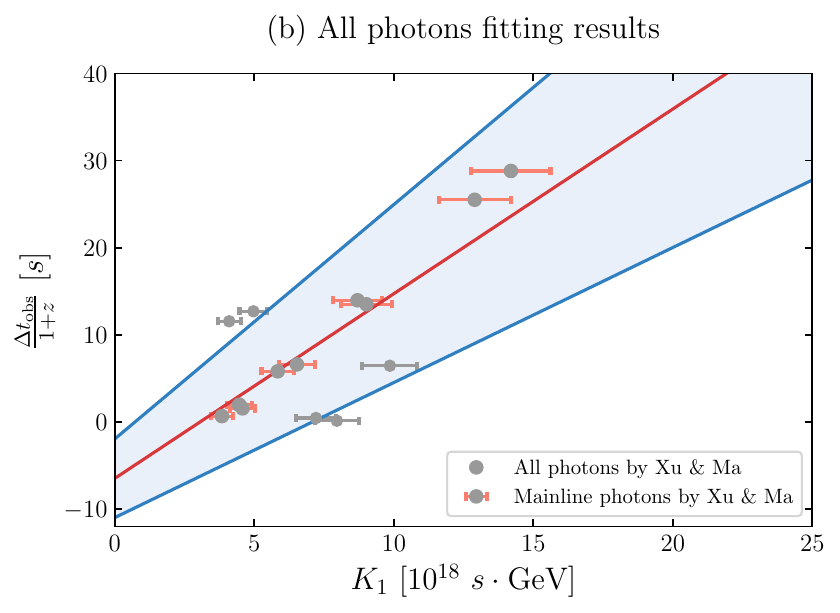}
    \end{minipage}
    \caption{Results of Model A using all 14 high-energy photons same as \cite{Xu:2016zxi, Xu:2016zsa}. The left panel shows the posterior distribution results, where the dashed lines of subfigures show the 1-$\sigma$ deviation from the central value. The right panel shows the fitting results for 14 high-energy photons. The gray dots with the error bars denote the same 14 high-energy photo events as \cite{Xu:2016zxi, Xu:2016zsa}. The gray dots with error bars in orange color denote the 9 mainline photons in   \cite{Xu:2016zxi, Xu:2016zsa}. The red line denotes the central value fitting results, and the two blue lines denote the 1-$\sigma$ deviation fitting results.}
    \label{model_A_1}
\end{figure}

\begin{figure}[htbp]
    \centering
    \begin{minipage}{0.43\textwidth}
        \centering
        \includegraphics[width=\textwidth]{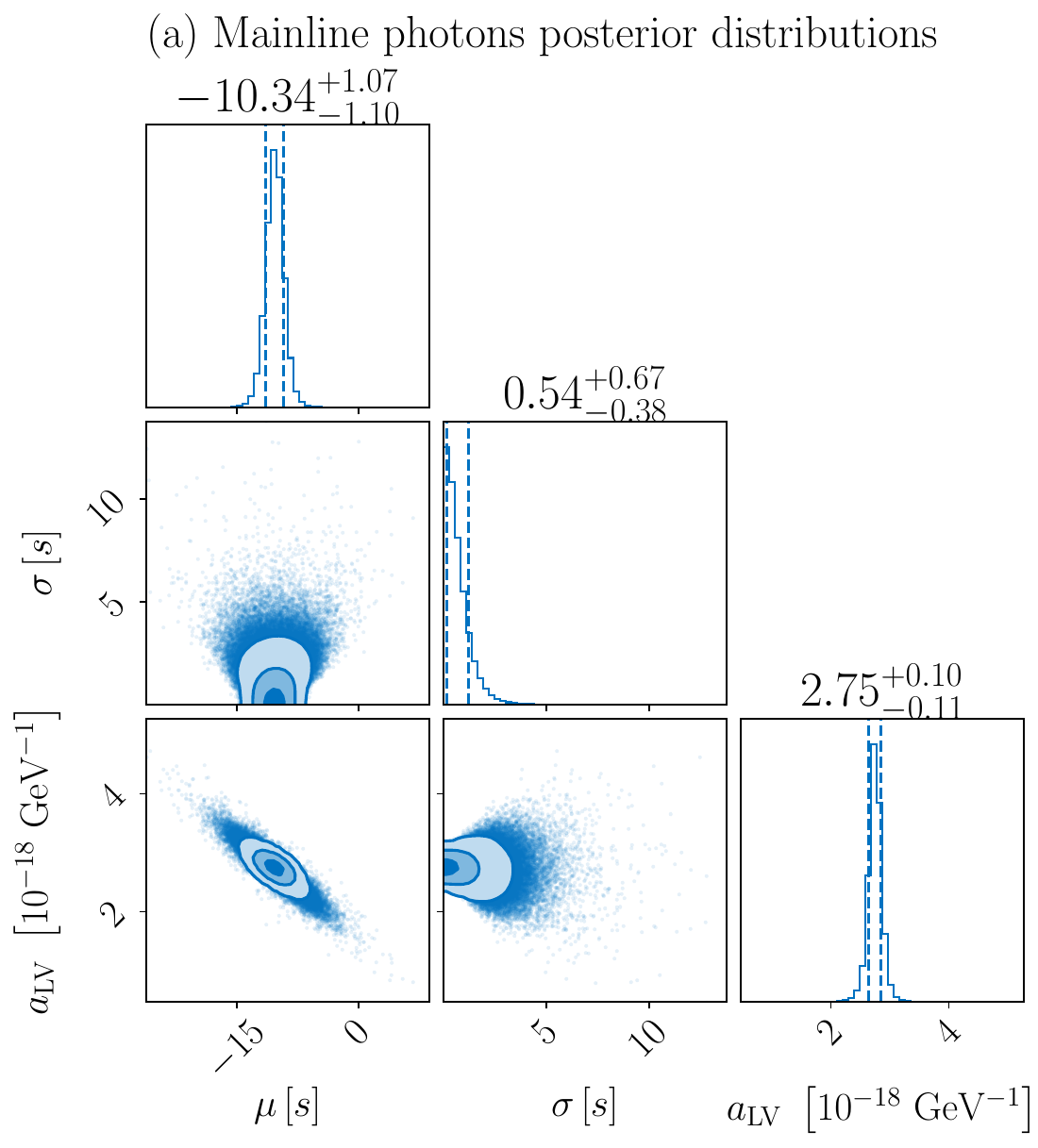}
    \end{minipage}
    \hfill
    \begin{minipage}{0.55\textwidth}
        \centering
        \includegraphics[width=\textwidth]{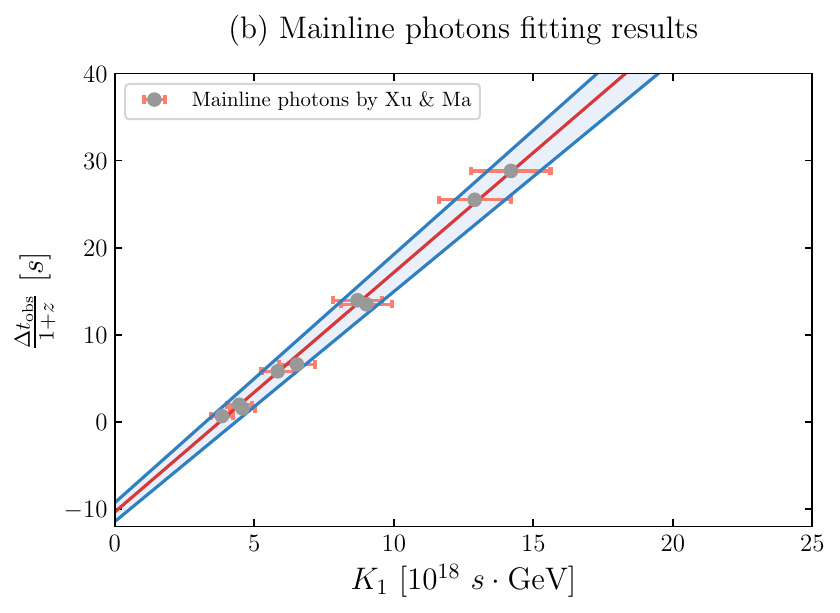}
    \end{minipage}
    \caption{Same as Fig.~\ref{model_A_1}, but for the 9 mainline high-energy photons as \cite{Xu:2016zxi, Xu:2016zsa}.} 
    \label{model_A_2}
\end{figure}

We use the same GRBs data as \cite{Xu:2016zxi, Xu:2016zsa}. As shown in Fig.~2 of \cite{Xu:2016zsa}, the distribution of those high-energy photons seems to show a regularity that they can be considered in two cases: the mainline photons case (9 high-energy photons are involved) and the all photons case (all 14 high-energy photons are involved). We also consider the same two cases in the discussion of Model A. The results for all photons case are shown in Fig.~\ref{model_A_1}. The left panel shows the posterior distributions and the right panel shows the fitting results where all 14 photons are denoted by gray dots and the 9 mainline photons are denoted by gray dots with orange error bars, respectively. The results suggest that the mean value of intrinsic time delay $\mu = -6.48^{+4.53}_{-4.50}$~s, and $a_{\rm LV} = 2.12^{+0.57}_{-0.57} \times 10^{-18} \ \rm{GeV^{-1}}$. The corresponding  $E_{\rm LV} = 4.71\textcolor{red}{^{+1.71}_{-0.99}} \times 10^{17}$~GeV\renewcommand{\thefootnote}{\ensuremath{\ast}}\footnote{There were misprints with errs of $E_{\rm LV}$ in the published version. The red color denotes the corrected values and the same for the rest corrections.}. 
The results for the mainline photons case are shown in Fig.~\ref{model_A_2}. The results suggest that the mean value of intrinsic time delay $\mu = -10.34^{+1.07}_{-1.10}$~s, and $a_{\rm LV} = 2.75^{+0.10}_{-0.11} \times 10^{-18} \ \rm{GeV^{-1}}$. The corresponding   $E_{\rm LV} = 3.64\textcolor{red}{^{+0.15}_{-0.13} }\times 10^{17}$ GeV.  Results for both two cases are consistent with the results in \cite{Xu:2016zxi, Xu:2016zsa}. Both cases suggest that high-energy photons are emitted earlier than low-energy photons because of the negative value of $\mu$, and the LV time delay term contributes to the observed time delay.

\subsection{Model B}
In principle, the regularity of observed time delay for GRBs might also be caused by other physical mechanisms. For example, high-energy photons ranging from 33.6 GeV to 146.6 GeV at the source frame might be emitted at different times. Therefore, in Model B, we assume that the observed time delay comes from the intrinsic time delay and neglect the LV time delay term. We assume the intrinsic time delay consists of two parts. One part is a common time delay term $\Delta t_{\mathrm{in,c}}$. The other part is an intrinsic time delay term correlated with the source frame energies of high-energy photons. For simplicity, we assume it is proportional to the energy value. At this situation, the observed time delay satisfies:
\begin{equation}
\frac{\Delta t_{\mathrm{obs}}}{1+z}= \alpha E_{s} + \Delta t_{\rm in,c},
\end{equation}
where $E_s$ is the source frame energy of high-energy photon and $\alpha$ is the coefficient. Again, we assume the common time delay term $\Delta t_{\mathrm{in,c}}$ follows Gaussian distribution and needs to be marginalized. The posterior for this model now is:
\begin{equation}
p \propto \exp\left[-\frac{1}{2}\sum_{j=1}^{p}\left(\frac{\left(\frac{\Delta t_{\mathrm{obs},j}}{1+z_{j}} -\mu - \alpha E_{s} \right)^{2}}{\sigma^{2}+ \alpha^2\sigma_{E_{s,j}}^2}+\ln(\sigma^{2} + \alpha^2\sigma_{E_{s,j}}^2)\right)\right]p\left(\mu\right)p\left(\sigma\right)p(\alpha).
\end{equation}

We also consider the priors of these parameters, $\mu$, $\sigma$, and $\alpha$, to be flat:
\begin{align}
    \begin{cases}
        p\left(\mu\right) \sim U\left[-30, 30 \right] \ {\rm s}, \\
        p\left(\sigma\right) \sim U\left[0, 30 \right] \ {\rm s}, \\
        p(\alpha) \sim U\left[-30, 30 \right] \ {\rm s \cdot GeV^{-1}}.
    \end{cases}
\end{align}

\begin{figure}[htbp]
    \centering
    \begin{minipage}{0.43\textwidth}
        \centering
        \includegraphics[width=\textwidth]{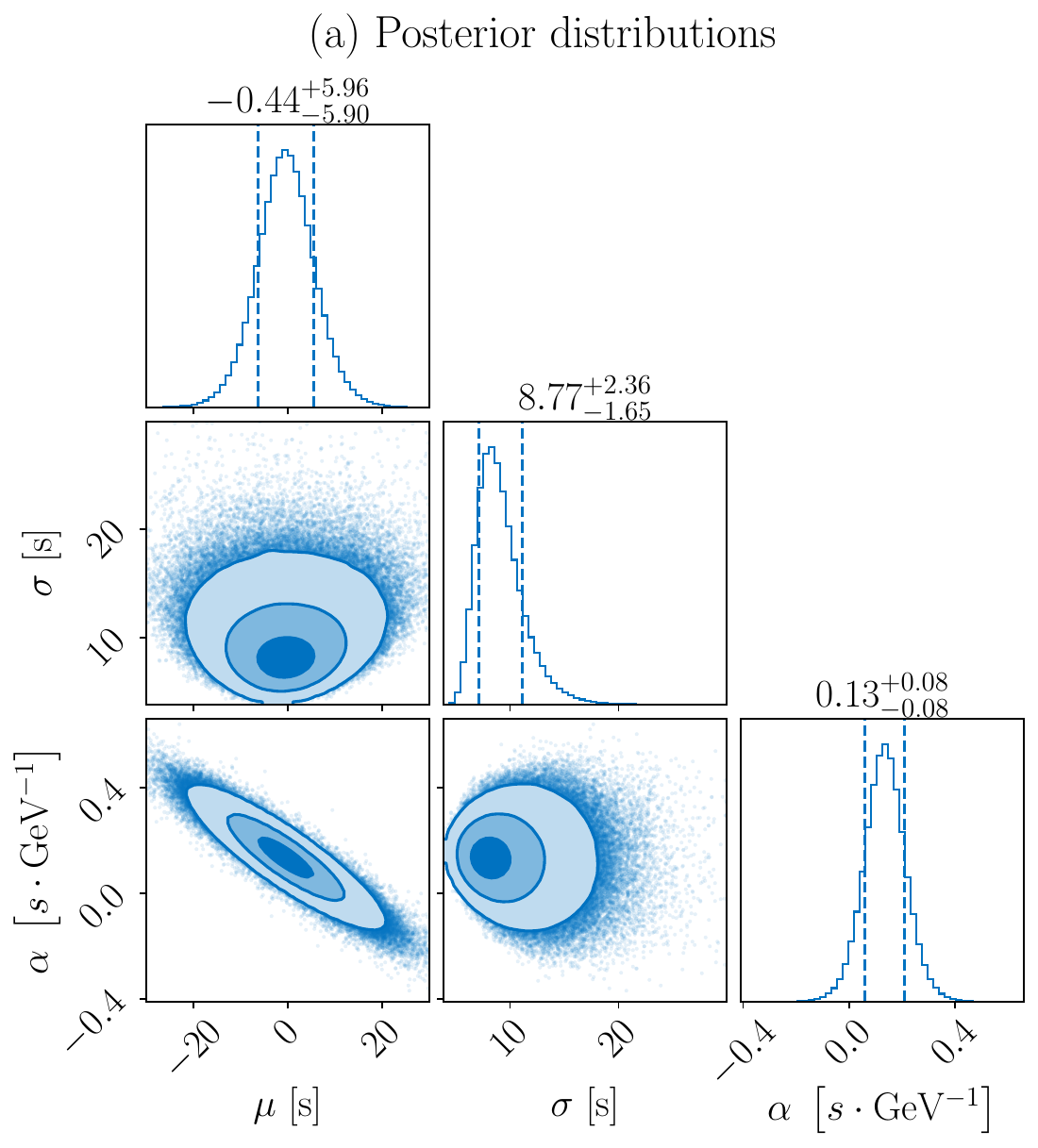}
    \end{minipage}
    \hfill
    \begin{minipage}{0.55\textwidth}
        \centering
        \includegraphics[width=\textwidth]{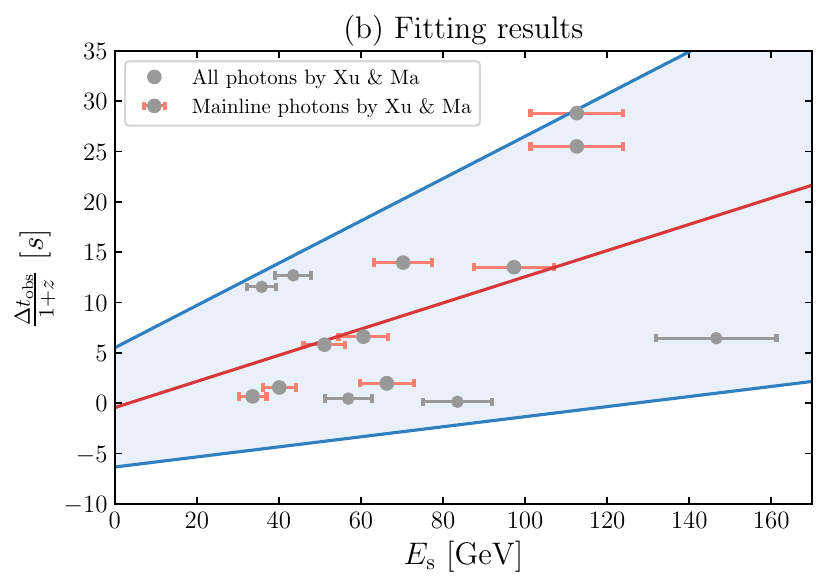}
    \end{minipage}
    \caption{Results of Model B for all 14 high-energy photons. The left panel shows the posterior distributions and the right panel shows the fitting results of Model B.}
    \label{model_B}
\end{figure}

Applying the same all 14 high-energy photons as Model A, we obtain the results shown in Fig.~\ref{model_B}. We do not divide the data into two cases in Model B. This is because the variables are changed from $K_1$ in Model A to $E_s$ in Model B . As shown in the right panel of Fig.~\ref{model_B}, there is no clear distinction between the 9 mainline high-energy photons in Model A (gray dotes denoted with orange error bars) and all 14 high-energy photons (all gray dotes). Therefore, we only calculate the posterior distributions for all photons in Model B. The results suggest that the mean value of intrinsic time delay is $\mu = -0.44^{+5.96}_{-5.90}$~s and the coefficient is $\alpha = 0.13^{+0.08}_{-0.08} \ { s \cdot \rm GeV^{-1}}$. The positive source energy correlated time delay term suggests that the high-energy photons are emitted later than the low-energy photons. With a tiny negative common intrinsic time delay term, Model B could also be a potential explanation for the observed time delay of GRB data.

\subsection{Model C}
For the high-energy photons from the same GRB,
we can not mathematically distinguish between the LV time delay term in Model A and the source energy correlated time delay term in Model B, although the physical mechanisms are 
totally different. However, when considering multiple sources of GRB events, we might distinguish those two terms because of their different responses to the redshifts of GRBs. Therefore, we combine the above two models into a unified model as Model C. We assume the observed time delay comes from two mechanisms: the LV time delay and intrinsic time delay with a common term plus a source energy correlated term. The observed time delay in this model now satisfies:
\begin{equation}
\frac{\Delta t_{\mathrm{obs}}}{1+z}=\Delta t_{\mathrm{in,c}} + \alpha E_{s} + a_{\rm LV}K_1 .
\end{equation}

After marginalizing 
over
the common time delay term $\Delta t_{\mathrm{in,c}}$, we obtain the posterior as:
\begin{equation}
p \propto \exp\left[-\frac{1}{2}\sum_{j=1}^{p}\left(\frac{\left(\frac{\Delta t_{\mathrm{obs},j}}{1+z_{j}} -\mu - \alpha E_{s,j} -a_{\mathrm{LV}}K_{j}\right)^{2}}{\sigma^{2}+ \alpha^2\sigma_{E_{s,j}}^2+a_{\mathrm{LV}}^{2}\sigma_{K_{j}}^{2}}+\ln(\sigma^{2}+ \alpha^2\sigma_{E_{s,j}}^2+a_{\mathrm{LV}}^{2}\sigma_{K_{j}}^{2})\right)\right]p\left(\mu\right)p\left(\sigma\right)p\left(a_{\rm LV}\right)p(\alpha).
\end{equation}

We consider the flat priors for the parameters $\mu$, $\sigma$, $a_{\rm LV}$, and $\alpha$:
\begin{align}
    \begin{cases}
        p\left(\mu\right) \sim U\left[-30, 30 \right] \ {\rm s}, \\
        p\left(\sigma\right) \sim U\left[0, 30 \right] \ {\rm s}, \\
        p\left(a_{\rm LV}\right) \sim U\left[0, 30 \right] \times 10^{-18} \ \rm{GeV^{-1}}, \\
        p(\alpha) \sim U \left[-30, 30 \right] \ {\rm s \cdot GeV^{-1}}. \\
    \end{cases}
\end{align}

\begin{figure}[htbp]
    \centering
    \begin{minipage}{0.45\textwidth}
        \centering
        \includegraphics[width=\textwidth]{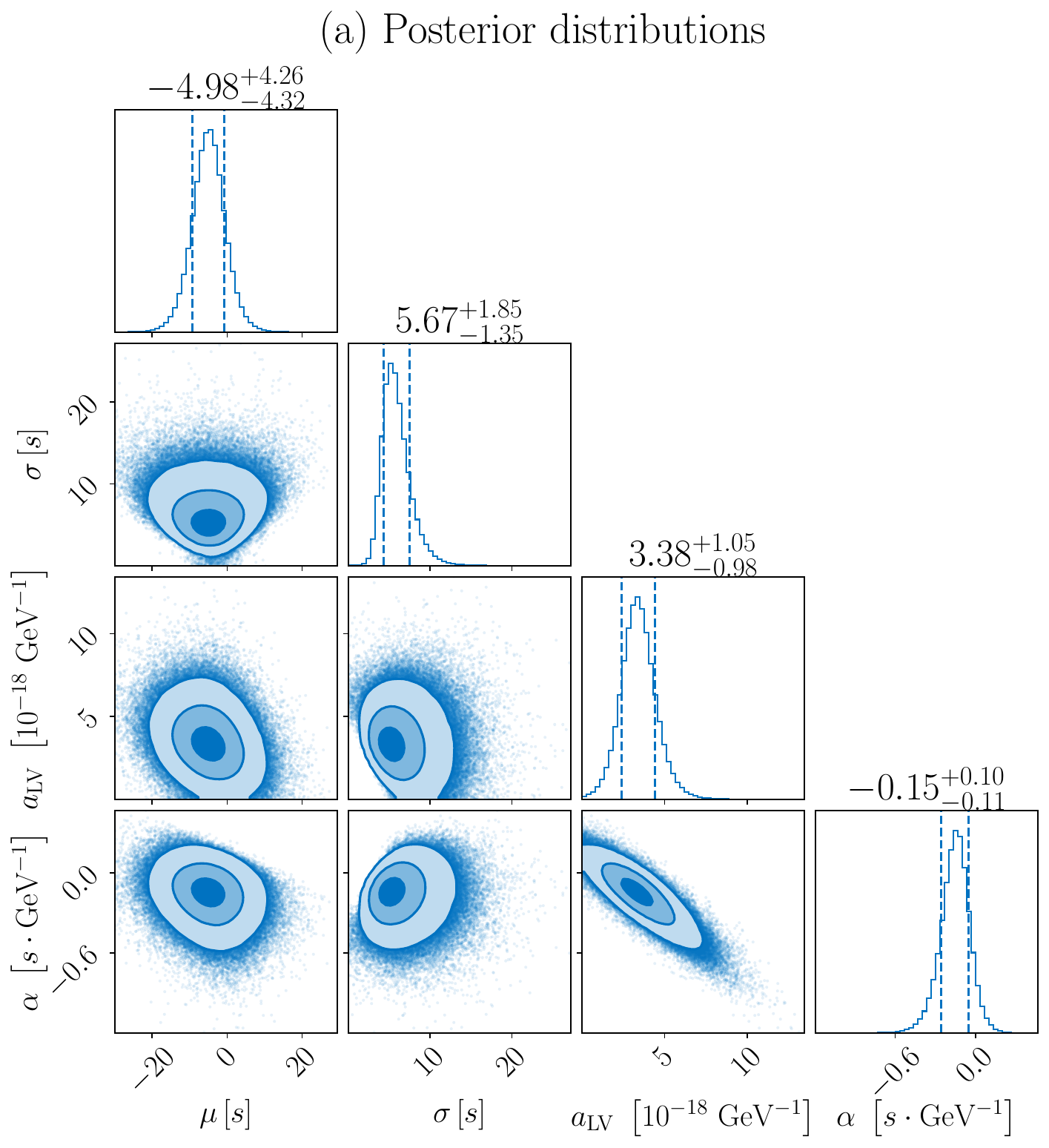}
    \end{minipage}
    \hfill
    \begin{minipage}{0.50\textwidth}
        \centering
        \includegraphics[width=\textwidth]{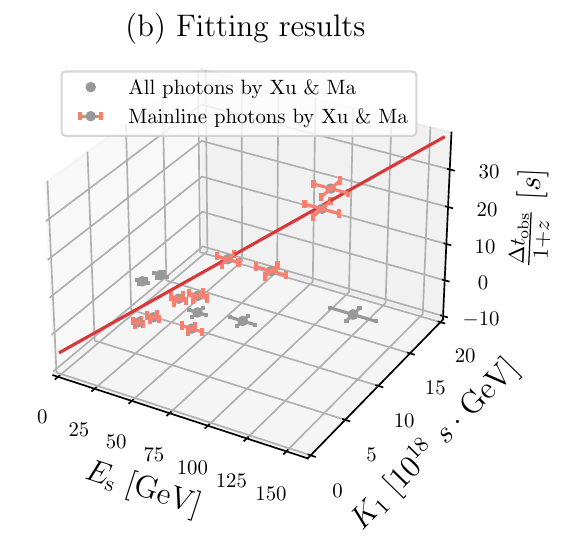}
    \end{minipage}
    \caption{Results of Model C for all 14 high-energy photons. The left panel shows the posterior distributions. The right panel shows the fitting results with the red line. The all 14 high-energy photons are denoted by gray dots with two-dimensional error bars in the $K_1$ axis and the $E_{\rm s}$ axis, respectively. The 9 mainline high-energy photons are denoted by gray dots with two-dimensional error bars in orange color.}
    \label{model_C}
\end{figure}

As there are both $E_s$-dependence and $K_1$-dependence in the fitting procedure rather than a simple linear relation, we should fit all 14 high-energy photons in Model C. 
Then we obtain the results shown in Fig.~\ref{model_C}. 
The left panel shows the posterior distributions and the right panel shows the fitting results. The results suggest that the mean value of intrinsic time delay $\mu = -4.98^{+4.26}_{-4.32}$ s, $a_{\rm LV} = 3.38^{+1.05}_{-0.98} \times 10^{-18} \ \rm{GeV^{-1}}$, and $\alpha = -0.15^{+0.10}_{-0.11} \ {\rm s \cdot GeV^{-1}}$. The corresponding  $E_{\rm LV} = 2.96\textcolor{red}{^{+1.21}_{-0.70}}\times 10^{17}$ GeV. We obtain a physics picture similar to that of Model A. Both the common intrinsic time delay term and the source energy correlated intrinsic time delay term are negative, which suggests that the high-energy photons 
are
emitted earlier than the low-energy photons. Meanwhile, the $E_{\rm LV} \simeq 2.96 \times 10^{17}$ GeV in Model C is smaller than the results in Model A as $E_{\rm LV} \simeq 4.71 \times 10^{17}$ GeV for all 14 high-energy photons case, implying a slightly stronger LV effect. It is worth mentioning that we still need more high-energy photons from GRBs to get more solid conclusions.

It is imperative to investigate the discrepancies in the constraints placed on $E_{\rm{LV}}$ in other studies compared to our findings. The primary factor contributing to these variations lies in the assumptions made regarding the timing of emission between high-energy and low-energy photons. For instance, shortly after the detection of a high-energy photon from GRB 090510 by Fermi LAT~\cite{FermiGBMLAT:2009nfe}, Ref.\cite{Xiao:2009xe} proposed a stringent constraint 
$E_{\rm{LV}}> 6.3 E_{\rm{P}}$ based on the assumption that high-energy photons cannot be inherently emitted before low-energy photons at the GRB source. Additionally, the analysis conducted by MAGIC~\cite{MAGIC:2020egb} on TeV scale photon events from GRB 190114C was carried out under the assumption that all photons were emitted after the trigger time $T_0$ at the source.

In contrast, our current analysis reveals a scenario where high-energy photons were emitted prior to low-energy photons at the GRB source solely through data fitting. This prediction of a preburst phase for high-energy photons has been discussed in previous studies~\cite{Zhu:2021pml,Chen:2019avc,Zhu:2021wtw} with favorable signals.

\section{Summary}
\label{summary}

In this letter, our primary focus is on investigating the energy-dependent intrinsic time delay effect and integrating it with the existing model to examine Lorentz invariance violation (LV) through fitting the Fermi Gamma-ray Space Telescope (FGST) high-energy photon data from Gamma-Ray Bursts (GRBs). We introduce a comprehensive Bayesian parameter estimation approach for a multiple linear model. Subsequently, we propose three distinct models to fit the high-energy photon data from GRBs as documented in previous studies \cite{Xu:2016zxi, Xu:2016zsa}.

Model A aligns with the approach adopted in prior research \cite{Xu:2016zxi, Xu:2016zsa, Zhu:2021pml}, yielding consistent outcomes. These findings indicate that high-energy photons are emitted prior to low-energy photons, with an associated Lorentz invariance violation energy scale of 
$E_{\rm LV} \simeq 4.71  \times 10^{17}$ GeV for the complete photon dataset.

In Model B, we exclude the consideration of LV effects and solely focus on intrinsic time delay. We postulate that the intrinsic time delay comprises two components: a shared time delay term 
$\Delta t_{\mathrm{in,c}}$
and an intrinsic time delay term proportional to the source frame energies of each high-energy photon. Unlike the scenario in Model A, where a distinction was made between the 9 primary high-energy photons and all 14 high-energy photons, in Model B, we incorporate the entire set of 14 high-energy photons for calculating the posterior distribution. The outcomes indicate a marginal negative common intrinsic time delay. With the inclusion of the positive source energy-correlated intrinsic time delay term, Model B emerges as a plausible explanation for the observed time delay phenomenon in GRBs.

Finally, in Model C, we integrate the two aforementioned models into a unified framework and perform fitting on the same set of 14 high-energy photons from GRBs as in Model A. The outcomes indicate the presence of a common negative intrinsic time delay term along with a negative source energy-correlated intrinsic time delay term. Consequently, a physical scenario akin to that of Model A emerges, revealing that high-energy photons are emitted prior to low-energy photons. Additionally, in comparison to the energy scale of Lorentz invariance violation (LV) in Model A, a reduced value of 
$E_{\rm LV} \simeq 2.96 \times 10^{17}$ GeV is obtained. This implies that even with the incorporation of the source energy-correlated intrinsic time delay, LV effects may still serve as a significant underlying physical mechanism for the observed time delay. It is important to note that the aforementioned conclusions are derived from the same dataset of GRB data as detailed in \cite{Xu:2016zxi, Xu:2016zsa}. Further GRB data are essential to distinguish and evaluate the respective contributions due to LV-induced time delays and intrinsic time delays, as well as to explore potential patterns in the arrival time data of high-energy photons from GRBs.
\\

\noindent
{\bf{Acknowledgements:}}
This work is supported  
by National Natural Science Foundation of China under Grants No.~12335006 and No.~12075003.
We thank Zhenwei Lyu, Jie Zhu, and Hao Li for helpful discussions and comments. This work is supported by High-performance Computing Platform of Peking University.

\bibliographystyle{elsarticle-num}
\bibliography{LV}

\end{document}